\documentclass[amsmat,amssymb,amsfonts,aps,prb,twocolumn,showpacs]{revtex4}

\usepackage{graphicx}
\usepackage{dcolumn}
\usepackage{bm}

\usepackage{xcolor}

\begin{document}

\title{Anisotropic intrinsic spin relaxation in graphene due to flexural distortions}

\author{S. Fratini (1), D. Gos\'albez-Mart\'inez (2) , P. Merodio C\'amara (1,3), J. Fern\'andez-Rossier (4)
  \footnote{Permanent Address: Departamento de Fisica Aplicada, Universidad de Alicante, Spain} 
  }
\affiliation{
(1) Institut N\'eel-CNRS and Universit\'e Joseph Fourier, Bo\^{\i}te Postale 166, F-38042 Grenoble Cedex 9, France
\\
(2) Departamento de F\'isica Aplicada, Universidad de Alicante, 03690
San Vicente del Raspeig, Spain \\
(3) SPINTEC, UMR CEA/CNRS/UJF-Grenoble 1/Grenoble-INP, INAC, Grenoble, France\\
(4)  International Iberian Nanotechnology Laboratory (INL),
Av. Mestre Jos\'e Veiga, 4715-330 Braga, Portugal}

\date{\today} 

\begin{abstract} 

We propose an intrinsic spin scattering mechanism in  graphene
originated by the interplay of atomic spin-orbit interaction and the
local curvature induced by flexural distortions of the atomic lattice.
Starting from a multiorbital tight-binding Hamiltonian with spin-orbit
coupling considered non-perturbatively, we derive an effective
Hamiltonian for the spin scattering of the Dirac electrons due to
flexural distortions. We compute the  spin lifetime due to both
flexural phonons and ripples and we find values in the  microsecond range at
room temperature.  Interestingly, this mechanism is anisotropic on two counts. First, the relaxation rate is different for off-plane and in-plane 
spin quantization axis.   Second, the spin relaxation rate depends on the angle formed by the crystal momentum with the carbon-carbon bond.  In addition, the spin lifetime is also valley dependent. 
The proposed mechanism  sets an upper limit for spin lifetimes in graphene and will be 
relevant when samples of high quality can be fabricated free of extrinsic sources of
spin relaxation. 

\end{abstract}

\maketitle

\section{Introduction.}
The electron spin  lifetime in carbon materials is expected to be very
long both because of the very large natural abundance of the spinless
nuclear isotope $^{12}C$ and the small size of spin orbit coupling. In
the case of flat graphene, the spin projection perpendicular to the
plane is conserved, even in the presence of the intrinsic spin-orbit
coupling.  Thus,  graphene was proposed as an optimal material to
store quantum information in the spin of confined
electrons \cite{Burkard07}. Most of the experiments 
\cite{Tombros-Nature,Tombros08,Kawakami1,Ozylmaz11,Kawakami2,Avsar11,Kawakami3,Folk2013,Maassen12,Beschoten2013}
show that the spin lifetimes are in the range of nanoseconds,  much shorter than expected from these
considerations, which is being  attributed to several 
 extrinsic factors: the breaking of reflection
symmetry due to coupling of graphene to a substrate \cite{Fabian09}
and/or a gate field,  the breaking of translational invariance due
to  impurities \cite{Neto09,Ochoa12,Mertig2013}, localized states\cite{Maassen13}, and  resonant coupling to extrinsic  magnetic moments\cite{Fabian13}.   
In the case of non-local spin valves,  the relaxation due the electronic coupling 
to  magnetic electrodes is also being considered\cite{Maassen12,Beschoten2013}. 
 
Here we take the opposite point of view and we consider    intrinsic spin relaxation in   graphene due
to the interplay between its  unique  mechanical and  electronic   properties.
 We show that flexural distortions, unavoidable in two dimensional
 crystals \cite{Fasolino2007}   in the form of either static ripples or
 out-of-plane  phonons,  induce  spin scattering between the  Dirac electrons, to
 linear order in the flexural field.  This coupling differs from the
 spin-conserving second-order interaction between Dirac
 electrons and flexural distortions \cite{Mariani09} that has been
 proposed as an intrinsic limit to mobility in suspended high quality
 samples \cite{limit1,limit2}.  

The fact that curvature enhances spin-orbit scattering  
has been  discussed\cite{Ando2000} and observed\cite{Kuemeth-Nature} in the case of carbon
nanotubes. Local curvature is also  expected to enhance  spin-orbit in
graphene \cite{Huertas06,Huertas09,Jeong2011,Dugaev} and
graphene ribbons \cite{Dani2011}.
In the  present paper 
we derive a microscopic Hamiltonian that 
describes explicitly the spin-flip scattering
of electronic states of graphene due to both dynamic and static 
flexural distortions. 
We describe graphene by means of a multi-orbital atomistic
description that naturally accounts for the two crucial ingredients of
the proposed intrinsic spin-phonon coupling: the intra-atomic spin
orbit coupling (SOC)  and the modulation of the inter-atomic integrals
due to atomic displacements.  Importantly, both the SOC  and the flexural
distortions couple the Dirac electrons to higher energy $\sigma$
bands,  in the spin-flip and spin conserving channels
respectively. Their  combined action results in  an effective
spin-flip interaction for the Dirac electrons. 

The spin-flip lifetimes 
 computed from our theory are in the range of $\mu s$   at room temperature. 
 Therefore, the observation of spin lifetimes in the nanosecond regime 
 implies that other extrinsic spin relaxation mechanisms are effective. 
 Our results provide an upper limit for the lifetime 
 that will be relevant when graphene samples   can be prepared without extrinsic sources of spin relaxation.  In the case of the proposed intrinsic spin relaxation mechanism, we  find that the spin liftetime of  electrons depends crucially on 
  the long wavelength mechanical properties of the  sample, 
 determined by its coupling to the environment.  We also find that the spin relaxation lifetime depends on the 
 quantization axis along which the spin scattering is taking place as well as on the relative angle between the electron crystal momentum $\vec{k}$ and the crystal lattice.

\section{Microscopic model.}
Our starting point is the tight-binding Hamiltonian 
${\cal H}={\cal H}_{SC} + {\cal H}_{SOC}$
  for electrons with spin $s$ moving in 
 a lattice of atoms $\vec{r}$, with atomic orbitals $o$.
We write the  hopping part as 
 \begin{equation}
\label{eq:SC}
 {\cal H}_{SC}= \sum_{\vec{r},\vec{r}',o,o',s}  
 H_{o,o'}( \vec{r}-\vec{r}') 
 \Psi^{\dagger}_{\vec{r},o,s} \Psi_{\vec{r}',o',s}.
 \end{equation}
considering explicitly the  dependence of the 
(spin conserving) inter-atomic matrix elements on the
positions of the atoms. 
 The intra-atomic  SOC reads
 \begin{equation}
{\cal H}_{SOC}=  \sum_{\vec{r},o,o',s,s'} \lambda \langle \vec{r} o \sigma|\vec{L}(\vec{r})\cdot\vec{S}|\vec{r}o's'\rangle
\Psi^{\dagger}_{\vec{r},o,s} \Psi_{\vec{r},o',s'}
 \end{equation}
where $\vec{S}$ is the spin operator,
$\vec{L}(\vec{r})$ is the orbital angular momentum operator acting upon the
atomic orbitals of site $\vec{r}$ and  $\lambda$ is the spin-orbit
coupling parameter.

Deviations from the ideal graphene lattice affect its electronic
properties via modifications of the
transfer integrals in Eq. (\ref{eq:SC}). Their
dependence on the inter-atomic distance, for example, 
gives rise to an electron-phonon interaction
analogous to that  of
conducting polymers \cite{SSH} and 
leads to 
the appearance of effective gauge fields \cite{Huertas09,RMP}. 
In the present model,  we consider instead 
the coupling with flexural distortions arising from the {\em angular}
dependence of the interatomic Hamiltonian.
We describe corrugations away from perfectly flat graphene in the form
$\vec{r}\simeq \vec{r}_0+ h(\vec{r}) \hat{z}$,
where $\vec{r}_0$ is a vector of the honeycomb lattice and
$h(\vec{r})$ is the displacement of atom $\vec{r}$ 
perpendicular to the graphene sheet.  
We expand the interatomic Hamiltonian matrix  
to lowest order in the flexural field
$h(\vec{r})$ and
rewrite  the Hamiltonian as ${\cal H}={\cal H}_0+{\cal V}$, where  
${\cal H}_0$ now describes ideally flat graphene 
including the weak intra-atomic SOC perturbation, and 
  \begin{equation}
 {\cal V}=  \sum_{r,r',o,o',s}\left[ h(\vec{r})-h(\vec{r}') \right]
 \frac{\partial}{\partial z} H_{o,o'}(\vec{r}_0-\vec{r}_0')
 \Psi^{\dagger}_{\vec{r},o,s} \Psi_{\vec{r}',o',s}.
  \label{e-p-atoms}
 \end{equation}
is the spin-conserving coupling between electrons and corrugations.

\section{Electron-flexural phonon scattering.}
\begin{figure}[h]
   \centering
\includegraphics[width=8.5cm]{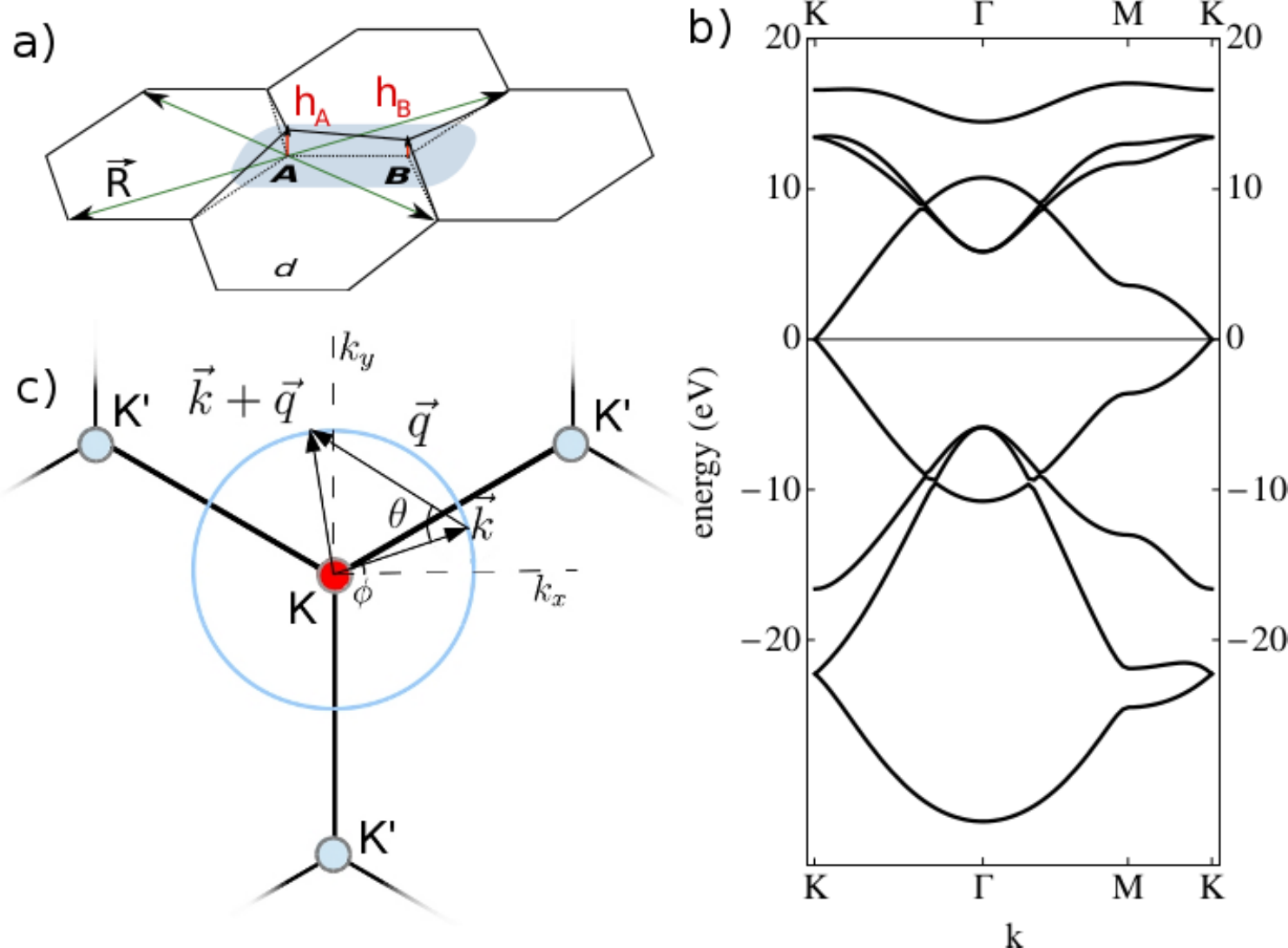}
   \caption{(color online) (a) Sketch of a local corrugation of the graphene layer. The shaded
     area is the unit cell. Green arrows indicate 
   the inter-atomic contributions to the
     electron-flexural phonon coupling of Eq. (\ref{Vkk}); 
(b) The band structure resulting from our
     Slater-Koster parametrization; (c)
   Kinematics of the scattering process around the Dirac point $K$ 
   in the low phonon frequency limit. }
 \label{fig:sketch}
 \end{figure}

 It is now convenient to recast Eq. (\ref{e-p-atoms}) in terms of the
 eigenstates of the Hamiltonian  ${\cal
   H}_0$ for flat graphene. 
   In this context, the atomic positions $\vec{r}=\vec{R}+\vec{r_\alpha}$, are specified
 by their unit cell vector, $\vec{R}$, and 
their position $\vec{r}_\alpha$ inside the cell (sublattice index 
$\alpha=A,B$). 
The creation operators for 
  Bloch states  are related to
 atomic orbitals through: 
\begin{equation}
 c_{\nu\vec{k}}^\dagger= 
\frac{1}{\sqrt{N}}\sum_{\vec{R},\alpha,o,s} e^{i\vec{k}\cdot\vec{R}} 
 {\cal C}_{\nu,\vec{k}}(\alpha,o,s) \Psi^\dagger_{\vec{R}+\vec{r}_\alpha,o,s} 
 \end{equation}
  where $\vec{k}$ is the wave-vector,  
the coefficients  ${\cal C}_{\nu,\vec{k}}(\alpha,o,s)$ are
  obtained from the diagonalization of the Bloch matrix and $\nu$ is
  an index that  labels the resulting bands 
(with mixed spin and angular momentum).
Similarly, we expand the  flexural field on each sublattice 
in its Fourier components,
 $ h_\alpha(\vec{R}) =\frac{1}{\sqrt{N}} \sum_{\vec{q}}
e^{-i\vec{q}\cdot\vec{R}}   
h_\alpha(\vec{q})$.
    After a lengthy but straightforward calculation, we can express
    Eq. (\ref{e-p-atoms}) as a term causing scattering between
    crystal states with different momentum and band indices: 
  \begin{eqnarray}
 {\cal V}= \sum_{\vec{k},\vec{k}',\nu,\nu'} {\cal V}_{\nu,\nu'}(\vec{k},\vec{k'})
   c^{\dagger}_{\nu,\vec{k}}c_{  \nu',\vec{k}' }
  \label{e-p-Bloch}
 \end{eqnarray}
  \begin{eqnarray}
 {\cal V}_{\nu,\nu'}(\vec{k},\vec{k'})
&  \equiv& 
    \frac{1}{ \sqrt{N}}
 \sum_{\vec{R},\alpha,\alpha',o,o',\sigma}
 \frac{\partial}{\partial z}
 H_{o,o'}(\vec{r}_\alpha-\vec{r}_{\alpha'}-\vec{R}) 
 \times \nonumber \\ &\times&
F^{\vec R}_{\alpha\alpha'}(\vec{k},\vec{k}')   {\cal C}^*_{\nu,\vec{k}}(\alpha,o,\sigma) 
   {\cal C}_{\nu',\vec{k}' }(\alpha',o',\sigma) 
   \label{Vkk}
 \end{eqnarray}
The coupling is {\em linear} in the flexural 
phonon field, through the form factor  
$F^{\vec R}_{\alpha\alpha'}(\vec{k},\vec{k}')= h_\alpha(\vec{k} -\vec{k}')e^{i\vec{k}'\cdot \vec{R}} -h_{\alpha'}(\vec{k}-\vec{k}')e^{i\vec{k}\cdot\vec{R}} 
 $.  
This should be contrasted with the  
electron-flexural phonon
coupling usually considered within the $\pi$ subspace  
\cite{RMP,Mariani09,limit1,limit2}, 
that is quadratic in the field
because of the quadratic dependence of interatomic distances on $h$.
Note that 
${H}_{oo'}$ is short ranged within our tight-binding description, 
which limits $\vec{R}$ to the
4 vectors connecting neighboring cells (inter-cell coupling), as
 illustrated in Fig. \ref{fig:sketch}a,
plus the null vector (intra-cell coupling).

\subsection{Slater-Koster parametrization.}
Eqs. (\ref{e-p-Bloch}) and (\ref{Vkk}) provide a general recipe  to
compute the coupling of electrons 
of a generic tight-binding Hamiltonian to a flexural field.
We now show that because we have included the SOC  in the reference Hamiltonian
${\cal H}_0$,
the perturbation Eq. (\ref{e-p-Bloch})  is able to
induce a direct coupling between
states with opposite spin.  
Following previous work \cite{Ando2000,Huertas06,Min06,Jeong2011,Dani2011} 
we consider a subset of  4 valence orbitals of the Carbon atom, namely
$o=s,p_x,p_y,p_z$, and  adopt a Slater-Koster (SK)
\cite{Slater-Koster} parametrization for the
tight-binding  Hamiltonian Eq. (\ref{eq:SC}) 
\cite{dorbitals}.
In this framework, the inter-atomic matrix elements $H_{o,o'}$ connecting the  atom
  $\alpha$ at $\vec{r}_{\alpha}$ with atom $\alpha'$ at
$\vec{r}_{\alpha'}+R$  can be expressed in terms of
4 parameters, $V_{ss}$, $V_{sp}$, $V_{\sigma\pi}$, $V_{\pi\pi}$,
describing inter-orbital overlaps in the $s,p$ basis 
\cite{Slater-Koster,Dani2011},  and
the three director cosines $l,m,n$ of the interatomic bond vectors
$\vec{\rho}=\vec{r}_\alpha-\vec{r}_{\alpha'}-\vec{R}$, 
defined by  $\vec{\rho}\equiv \rho  \left(l\hat{x}+m\hat{y}+n\hat{z}\right)$.  
Setting $U_{x}\equiv V_{pp\pi}+ x^2 V_{\sigma\pi}$ and
$V_{\sigma\pi}\equiv (V_{pp\sigma}-V_{pp\pi})$ we can write the SK
matrix  in
compact form as
\begin{equation}
H(\vec{\rho})=\left( \begin{array}{cccc} 
V_{ss\sigma} &l V_{sp\sigma} & m V_{sp\sigma} &nV_{sp\sigma} \\
-l V_{sp\sigma} & U_l  & lm  V_{\sigma\pi} & ln V_{\sigma\pi} \\
-m V_{sp\sigma} & lm  V_{\sigma\pi} & U_m &  mn V_{\sigma\pi} \\
-n V_{sp\sigma}& ln  V_{\sigma\pi}  & mn V_{\sigma\pi} &U_n  \end{array}\right)
\label{SKmatrix}
\end{equation}

The unperturbed crystal states for flat graphene  are described by
$H(\vec{\rho})$ with $n=0$ (all bonds within the $x,y$ plane). The
resulting band structure is shown in Fig. \ref{fig:sketch}b. 
From Eq. (\ref{Vkk}), the electron-flexural distortion coupling
is determined by  $\partial_z
H(\vec{\rho})=(1/d)\partial_nH(\vec{\rho})$,
where $d$ is the equilibrium C-C distance. 
Direct inspection of Eq. (\ref{SKmatrix}) shows that   
$H(\vec{\rho})(n=0)$ does not couple the $p_z$ and  $s,p_x,p_y$  sectors,
   while $\partial_n H(\vec{\rho})$ does.   
   As a result,  the $\pi$ and $\sigma$ bands of flat graphene 
    are   not mixed in the  absence of SOC, 
unless they scatter with flexural distortions. 
      In the presence of SOC, however,  the low-energy $\pi$ states
    with spin $s$ mix with the  $\sigma$ states with opposite spin
    already within the reference ${\cal H}_0$ for  flat graphene.
    Close to the Dirac points, where the $\pi$ and $\sigma$ bands are
    separated in energy by a gap $E_{\sigma\pi}$, the spin
    $\pi-\sigma$ mixing is   proportional to
    $\lambda/E_{\sigma\pi}$.  
     Since this correction is small,  
 the low energy Dirac bands of ${\cal H}_0$  can still be labeled
according to their dominant spin
character, that we denote as $\Uparrow$ and $\Downarrow$.   


The above derivation shows that there are  
two perturbations that couple $\pi$ and $\sigma$
states, the spin-conserving coupling to the flexural field, and the
spin-flip SOC. Their combination is able to yield a spin-flip channel
within the low energy $\pi$ bands, that is linear in both the flexural
deformation and in the atomic spin-orbit coupling $\lambda$. This is  very 
similar to the so called
Rashba spin orbit coupling, 
induced by the combination of  $\pi-\sigma$ mixing due
 to an external electric field and atomic
 spin-orbit coupling 
 \cite{Min06,Huertas06,Dugaev}, and 
different from  the $\lambda^2$ scaling of the intrinsic SOC
in flat graphene.

\subsection{Effective  Hamiltonian}
We now apply the microscopic theory developed above to obtain
the effective spin-flip Hamiltonian  
for electrons 
close to the Dirac points.
Anticipating that the 
dominant contributions to spin-flip scattering arise from the
long-wavelength, \textit{i.e.} low energy flexural modes, we  
consider the lower flexural branch for which $\omega_q\propto
q^2$ (see below), 
discarding the higher energy
modes that involve  out of  phase vibrations of the two sublattices.
 The flexural field is factored out from Eq. (\ref{e-p-Bloch}) by
 setting  $h_A(\vec{q})=h_{\vec{q}}$ and 
 $h_B(\vec{q})=
 e^{i\vec{q}\cdot(\vec{r}_B-\vec{r}_A)}
h_{\vec{q}}$, which yields:
  \begin{eqnarray}
 {\cal V}= \sum_{\vec{k},\vec{q},\nu,\nu'}
 {M}_{\nu,\nu'}(\vec{k},\vec{q}) \frac{h_{\vec{q}}}{d}
   c^{\dagger}_{\nu,\vec{k}+\vec{q}}c_{  \nu',\vec{k} }
  \label{e-p-Bloch2}
 \end{eqnarray}
with $ {\cal V}_{\nu,\nu'}(\vec{k}+\vec{q},\vec{k}) = 
\frac{h_{\vec{q}}}{d} M_{\nu,\nu'}(\vec{k},\vec{q})$.
The standard  form for the phonon spin-flip
interaction in second quantization is readily obtained by
substituting  $h_{\vec{q}}= \sqrt{\frac{\hbar}{2 M_C \omega_{
      \vec{q}}}}(a^{\dagger}_{-\vec{q}}+ a_{\vec{q}})$ 
 into Eq. (\ref{e-p-Bloch2}), with $M_C$ the Carbon mass.

\subsection{Spin quantization axis}

In the case of systems with  time reversal invariance and  inversion symmetry,  a Bloch state
 with momentum $\vec{k}$
has a twofold   Kramers degeneracy. This is  definitely the case of the ideal flat graphene. 
 As a result, there are infinitely many possible choices of the pairs of degenerate 
 states $\nu$ and $\nu'$.  In the calculations below we select a given pair by 
 including in the Hamiltonian an external magnetic field along the direction $\hat{n}$
 with magnitude negligible compared with all other energy scales in the problem, but enough to split the Kramers doublet and choose its spin quantization axis. 
 Importantly, the effective electron-phonon coupling depends on this choice, {\it i.e.}, it depends on $\hat{n}$.
 In the following we include $\hat{n}$ as an argument of the  phonon 
 spin-flip coupling and we label the two bands
 as $\Uparrow$ and $\Downarrow$, which are referred to the quantization axis $\hat{n}$.
  The fact that the phonon spin-flip coupling depends on $\hat{n}$   means that the 
  strength of the spin-flip Hamiltonian is not isotropic in the spin space.  This will lead 
  to an  anisotropy of the spin relaxation in graphene, closely
  related to the one recently proposed in the case of metals\cite{Zimmerman-2012}.

\begin{figure}[h]
  \centering
\includegraphics[width=6.0cm]{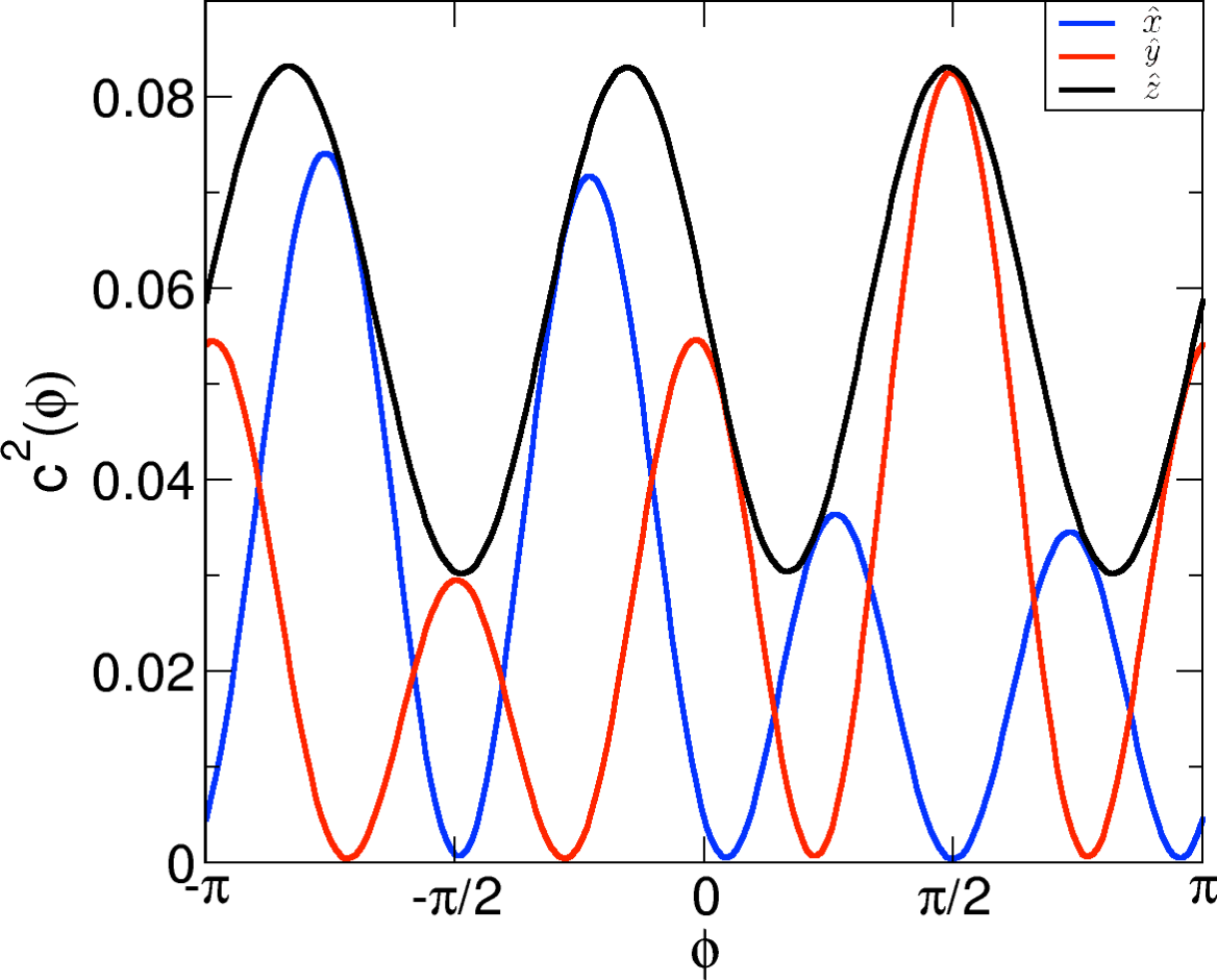}
  \caption{(color online) Spin-flip matrix elements as a function of the
  initial state wave-vector orientation $\phi$, for three different 
  orientations of the spin 
  quantization axis, $\hat{n}= \hat{x},\hat{y},\hat{z}$. It can be checked that  
$c^2(\phi,\hat{z})=c^2(\phi,\hat{x})+ c^2(\phi,\hat{y})$. 
}
  \label{fig:M}
\end{figure}

\section{Spin relaxation}

\subsection{Spin relaxation Rates}
The spin relaxation rate can now be calculated from  Eq. (\ref{e-p-Bloch2}) via the Fermi golden rule.
  Because the dispersion of the flexural modes is much weaker than the  electronic dispersion,
we can safely neglect the phonon frequency in the energy conservation.
The relaxation rate for an electron with momentum $\vec{k}$ in the band
  $\Uparrow$ is then obtained by summing over both phonon absorption and
  emission processes and over all possible final states in the
  $\Downarrow$ band, which yields:
\begin{equation}
  \label{eq:rate}
  \Gamma_{\vec{k},\hat{n}}=\frac{2\pi}{\hbar}\int \frac{d^2q}{(2\pi)^2}
|M_{\Uparrow,\Downarrow}(\vec{k},\vec{q},\hat{n})|^2 \langle h_{\vec{q}}^2 \rangle
\delta(E_{\vec{k}+\vec{q}}-E_{\vec{k}}). 
\end{equation}
This,  together with the explicit expressions for the  spin-flip matrix elements in Eq. (\ref{Vkk}), 
constitutes the main result of this work.
From Eq.  (\ref{eq:rate}) it is clear that 
once the specific form of the electron-flexural phonon coupling,
$M_{\Uparrow,\Downarrow}(\vec{k},\vec{q},\hat{n})$, is known,
the behavior of the spin relaxation rate is fully determined  by the statistical 
fluctuations of the flexural field, $\langle h_{\vec{q}}^2 \rangle$. 
Interestingly, the above expression describes on an equal footing 
both low-frequency 
flexural modes (that arise in free-standing  or weakly bound graphene
or graphite) 
as well as static ripples (relevant to graphene 
deposited on a substrate). The proposed
spin relaxation mechanism therefore applies without distinction to both
physical situations.

For actual calculations we   approximate the $\pi$ band energies as 
$E_{\vec{k}}=\pm \hbar v_F k$, which is
valid except for a negligible interval
around the Dirac point, where the SOC opens a gap of the order of few
$\mu eV$.   Energy conservation implies
$ \sqrt{k^2+q^2+2kq\cos\theta}=\sqrt{k^2}$.
This fixes the relative angle between $\vec{k}$ and $\vec{q}$ 
to $\cos \theta=-\frac{q}{2k}$, cf. Fig. \ref{fig:sketch}c. 
This equation has two solutions, that we label with the index $s$, 
which allows to perform the angular
integration in Eq. (\ref{eq:rate}), yielding:
\begin{eqnarray}
  \label{eq:q-integral}
  \Gamma_{\vec{k},\hat{n}}&=&
  \sum_{s}
  \frac{1}{\pi \hbar}\int_0^{2k} dq
\frac{ |M^{s}_{\Uparrow,\Downarrow}(k,q,\hat{n})|^2 \langle h_q^2 \rangle
  }{\hbar v_F\sqrt{1-(q/2k)^2}}.
\end{eqnarray}
We see that only long-wavelength 
fluctuations with $q\le 2k$  contribute to the spin relaxation.
From our numerics, we find that in the relevant case of
 small $k$ and $q\to 0$, 
the matrix elements evaluated on the energy-conserving surface (i.e. on shell, 
where $q/2k=-\cos \theta$)
satisfy:
\begin{equation}
  \label{eq:f}
  \sum_s|M^s_{\Uparrow,\Downarrow}(k,\hat{n})|^2 
\approx  c^2(\phi,\hat{n})  \lambda^2 q^4 d^4,
\end{equation}
where $\phi$ is the  angle formed between $\vec{k}$ and the $x$-axis in reciprocal space
and $c(\phi,\hat{n})$ is a dimensionless coefficient that  
only depends on the angle $\phi$, the spin quantization axis and the SK parameters. 
We plot this coefficient in Fig. 2, for 3 different orientations of the
spin quantization axis.

It must be noted that, 
 while Eq. (\ref{eq:f}) is strictly valid in the limit $q\ll 2k$,
it constitutes the dominant momentum dependence
of the matrix elements in the whole
integration range $q\le 2k$. Therefore,  
the angular dependence of $c(\phi)$ goes a long way to account for the 
$\phi$ dependence of the
 spin relaxation rate  that we discuss below.

\subsection{Fluctuations of the flexural field}
The scattering rate Eq. (\ref{eq:q-integral}) depends on the statistical fluctuations of the flexural field 
$\langle h_{\vec{q}}^2\rangle$, which we evaluate here for different scenarios.

We start with the  expression for  free-standing graphene at thermal
equilibrium: 
\begin{equation}
  \label{eq:fluct}
  \langle h_q^2 \rangle
  =\frac{\hbar}{2M_C\omega_q}[1+2n_B(\omega_q)]\simeq \frac{ k_BT}{M_C \omega_q^2},
\end{equation}
where $n_B(\omega_q)$ is the thermal population of mode $q$ and 
the second equality holds when $k_BT\gg\hbar \omega_q$.
For purely harmonic 
flexural modes, 
for which $\omega_q\simeq  D q^2$, the fluctuation
$\langle h_{\vec{q}}^2\rangle$  diverges as $q^{-4}$ for  small $q$.  When inserted into 
Eq. (\ref{eq:q-integral}), this divergence exactly compensates the $q^4$ dependence 
of the matrix element Eq. (\ref{eq:f}). In real samples, however, the singularity
of low-wavelength fluctuations is renormalized due to the 
interaction with other phonons (i.e. by anharmonic
effects) \cite{Zakahr10}, and can be 
further cut off by the presence of strain \cite{limit2} or
pinning to a substrate \cite{Sabio}.
The resulting dispersion can be parametrized as   
$\omega_q = D\sqrt{q^4+q^{4-\eta}q_c^{\eta}}$ for $\eta>0$ so that,
in the long wavelength limit \cite{Zakahr10,KatsGeim}, 
\begin{equation}
  \label{eq:hq}
  \langle h_q^2 \rangle \propto \frac{1}{q^{4-\eta}q_c^{\eta}}
\end{equation}
where $\eta$ and $q_c$  depend on the physical mechanism of
renormalization. Specifically, substrate pinning 
opens a gap in the phonon spectrum  \cite{Sabio},
corresponding to $\eta=4$; strain makes the dispersion
linear at long wavelengths \cite{limit2} ($\eta=2$);  anharmonic
effects yield $\eta=0.82$ 
\cite{Zakahr10}.   Substrate roughness also gives rise to fluctuations in the form
of Eq. (\ref{eq:hq}),
with $\eta=1$ \cite{Ishigami,KatsGeim}.

\section{Results and discussion.}

\subsection{Approximate estimate of the rate}
Before we discuss the results of our numerical integration of Eq. (\ref{eq:q-integral}) 
it is convenient to obtain an approximate analytical formula from the  integration
of the small $q$ part. 
   For that matter, we make use of the long wavelength expression Eq. (\ref{eq:f}),
drop the square root factor in the denominator, 
which is only relevant in a very narrow region around the backscattering 
condition $q\approx 2k$, and use the asymptotic expression Eq. (\ref{eq:hq})
for flexural fluctuations.  
The approximate expression for the rate reads:
\begin{eqnarray}
  \label{aprox}
\Gamma_{\vec{k}_F,\hat{n}} \simeq \frac{d}{\pi \hbar}  
\frac{\lambda^2  c(\phi,\hat{n},\tau)}{\hbar v_F}     \frac{(2k_Fd)^{\eta+1}}{(\eta+1) 
(q_c d)^\eta} 
r^2(T),
  \label{an0}
  \end{eqnarray}
which is valid at densities such that $k_F\ll q_c$.
Here we have  defined $r^2(T)=\frac{k_BT d^{2}}{M_CD^2} \frac{1}{d^2}$, 
representing the ratio 
between the short-range flexural fluctuations (i.e. Eq. (\ref{eq:fluct}) evaluated 
at $q= 1/d$) and the interatomic distance $d$.
From Eqs. (\ref{eq:fluct}) and (\ref{aprox}) we see that in ideal graphene with $\eta=0$ 
 the spin relaxation rate
 increases linearly with temperature, following the thermal population
 of flexural phonons. A weaker temperature dependence 
arises when anharmonic effects dominate, because  the
 anharmonic cutoff is itself temperature dependent, $q_c\propto
 \sqrt{T}$  \cite{Fasolino2007,Zakahr10}. 
 
   Eq. (\ref{an0}) permits a quick estimate of the efficiency of the spin rate.  
   We see that for a given value of $q_c$, the spin lifetime increases 
 as the exponent $\eta$ increases.
   A lower limit for the relaxation time is therefore 
   obtained by setting $\eta=0$ which, for
 $\lambda=8$meV,  $c=10^{-1}$,   
 $v_F=10^{-6}ms^{-1}$ and    $r^2\simeq10^{-2}$, yields a lifetime
 $\tau_s=1/\Gamma_{\vec{k}_F}$  on the order of  $1\mu$s at a density 
 $n=10^{12}cm^{-2}$ and at room temperature.  
 Lifetimes in the $\mu$s range are also obtained in the case of
 static ripples arising from the roughness of the underlying substrate, as we have  
 checked using the 
 values of $r^2$ and $q_c$ deduced from the height profiles in Ref. \cite{Ishigami}.
 In that case the lifetime is temperature independent.

\subsection{Energy dependence of the intrinsic spin  relaxation }
\begin{figure}[hbt]
  \centering
\includegraphics[width=4.0cm]{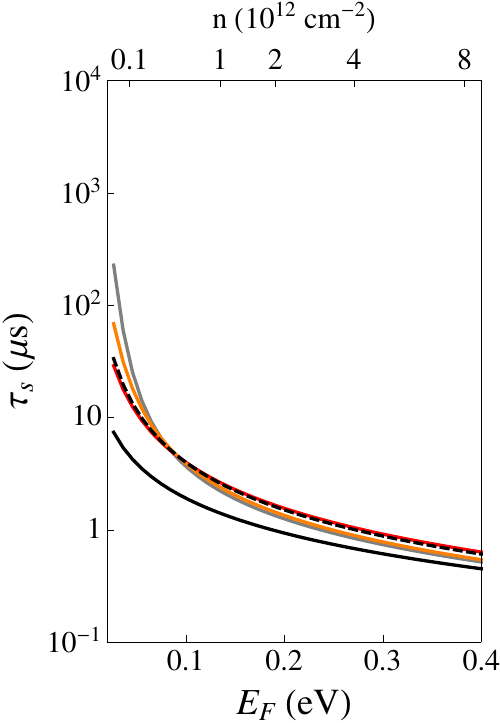}
\includegraphics[width=4.0cm]{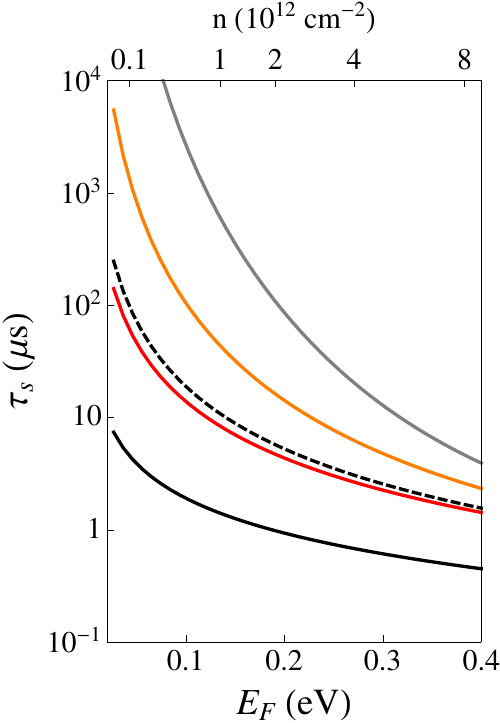}
  \caption{(color online) Room temperature  
spin lifetime calculated  for 
electrons at the Fermi energy with momentum parallel to the $x$-axis ($\phi=0$), 
for 2 different values of cutoff
momentum: (a) $q_c=0.01$ \AA $^{-1}$  and (b) $q_c=0.1$  \AA $^{-1}$
    (right) and  different long-wavelength scaling laws:  
   ideal graphene    (black, $\eta=0$);  including anharmonic effects
   (red, $\eta= 0.82$); including strain
    (orange, $\eta=2 $) and substrate pinning (gray, $\eta=4
    $). The black dashed line is for
    $\eta=1$, which is representative for substrate roughness. }
  \label{fig:rates1}
\end{figure}

We now compute Eq. (\ref{eq:q-integral}) numerically, without  analytical approximations. 
The spin lifetime $\tau_s=1/\Gamma_{k_F,\hat{n}}$ 
so obtained for electrons at the Fermi level is plotted in Figs.  
\ref{fig:rates1} and  \ref{anis-1}.

In Fig. \ref{fig:rates1} we show the  the spin lifetime
as a function of $E_F=\hbar v_F k$, 
fixing the momentum direction $\phi=0$, the valley $K$ (cf. Fig. 1c), 
and taking as spin quantization axis
the off-plane direction $\hat{n}=\hat{z}$.   We take \cite{limit2} 
$D=4.6 \cdot 10^{-7} m^2 s^{-1}$, $\lambda=8$
meV \cite{Kuemeth-Nature},  
 $T=300K$ and $v_F=1.16\cdot 10^6 m/s$ from our SK band structure.
In each panel of Fig. \ref{fig:rates1}, different curves correspond to different 
values of the scaling
exponent $\eta$, i.e. to physically different mechanical environments
for graphene.   Panels (a) and (b) correspond to  two different values of the cut-off 
momentum $q_c$. 
Two representative values,
(a) $q_c=0.01$ \AA $^{-1}$  and (b) $q_c=0.1$  \AA $^{-1}$ are considered,
 covering the large spread
of $q_c$ values available in the literature. 
In both panels,
the result for ideal graphene in the harmonic approximation 
is shown for reference (black) as it provides an absolute 
lower bound to the actual lifetime, in agreement with the estimate $\tau_s\sim 1\mu$s 
given after Eq. (\ref{an0}). 
Comparing Figs. \ref{fig:rates1}a and \ref{fig:rates1}b 
we see that the effect of the mechanical environment becomes more pronounced
for large values of the cutoff momentum $q_c$.
Because the spin relaxation is dominated by
the low energy fluctuations of the membrane, the  shortest 
spin lifetimes, excluding the harmonic theory,  
are obtained in suspended unstrained graphene (red curve),
i.e.  when   external mechanical influences are minimized and the mobility is possibly
largest. Even longer spin lifetimes can in principle be achieved by suppressing the 
fluctuations of the graphene membrane, by an applied strain (orange) or substrate pinning (gray). 
Pinning by interlayer binding forces should also inhibit the spin relaxation in epitaxially 
grown graphene.

For all the mechanical models considered here,  the spin relaxation time is 
a decreasing function of the density,  because the 
phase space for spin-flip scattering increases with $k_F$.
The density dependence within the different models can 
be anticipated by substituting the Fermi wavevector  $k_F=\sqrt{\pi n}$   
in the analytical expression Eq. (\ref{an0}), which 
results in
$\tau_s \propto n^{-(\eta+1)/2}$  for $k_F\lesssim q_c$
\cite{kf}.

Finally, it is apparent that in all instances the computed 
lifetimes are larger than 500$ns$. 
Therefore, the proposed intrinsic spin relaxation mechanism cannot account 
for present experimental observations where the spin lifetime is in the nanosecond range, 
which are presumably dominated by other (extrinsic) relaxation mechanisms.

\subsection{Anisotropy}

\begin{figure}
\centering
\begin{tabular}{cc}
\includegraphics[width=0.45\linewidth]{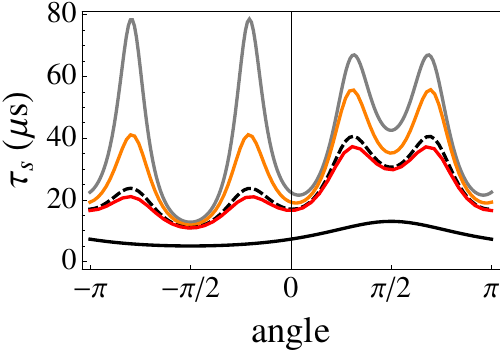} & 
\includegraphics[width=0.45\linewidth]{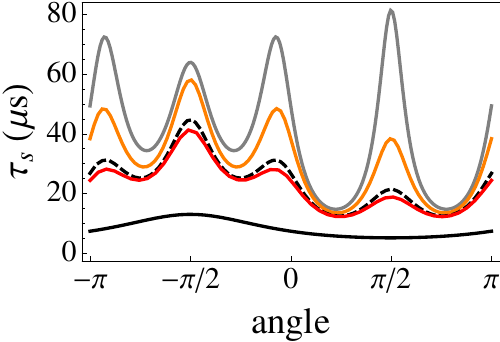} \\ 
\includegraphics[width=0.45\linewidth]{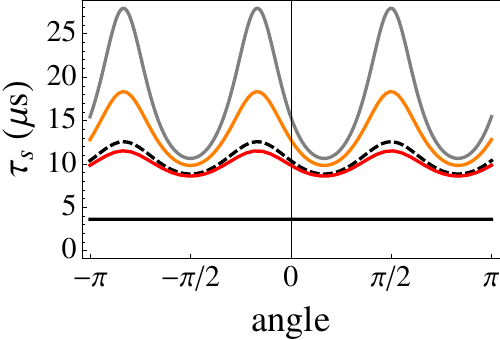} &   
\includegraphics[width=0.45\linewidth]{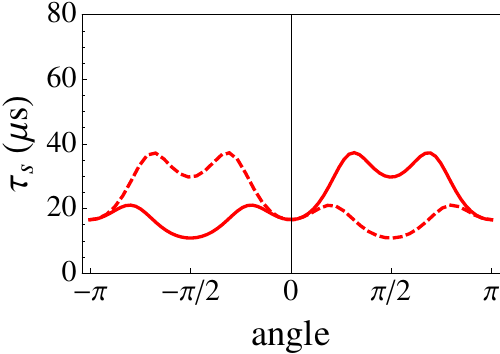} 
\end{tabular}
 \caption{(color online) Spin lifetime as a function of $\phi$, the polar angle of the 
 momentum of the initial state, 
at  room temperature,   calculated  for $E_F=54$meV, momentum cutoff $q_c=0.01\AA^{-1}$,  
for quantization axis $\hat{n}=\hat{x}$ (a) , 
$\hat{y}$ (b) and $\hat{z}$ (c),   and  different long-wavelength scaling laws:  
   ideal graphene    (black, $\eta=0$);  including anharmonic effects
   (red, $\eta= 0.82$); including strain
    (orange, $\eta=2 $) and substrate pinning (gray, $\eta=4 $). The black dashed line is for
    $\eta=1$, which is representative for substrate roughness.  
    Panel (d):  showing the results for the anharmonic case $\eta=0.82$, 
    quantization axis $\hat{n}=\hat{x}$ (i.e same as in (a))
    with the initial state in two different valleys.} 
    \label{anis-1}
\end{figure}

We now consider the influence of the momentum orientation, the spin quantization axis 
and the valley on the spin relaxation time.  The results of Fig. \ref{fig:rates1} 
have been obtained for 
$\phi=0$, $\hat{n}=\hat{z}$ and  valley K (cf. Fig. 1c).
We find that the spin relaxation lifetime of a state with momentum $\vec{k}$ depends on the 
angle $\phi$ formed between $\vec{k}$ and  the $\hat{x}$ direction,  
the spin quantization axis, and the valley. 
 Results for the angular and valley dependence are 
 shown in Fig. \ref{anis-1} for  $E_F=54$meV. Let us consider first $\hat{n}=\hat{z}$, 
 Fig. \ref{anis-1}c.  
 In this case the $\phi$ dependence shows  
 $C_3$ rotation symmetry, dephased with respect to the one of the lattice.  
 The momentum-direction anisotropy is not a 
 small effect, as the lifetime changes by more than a factor  2  between maxima and minima,  
 for both values of $E_F$.  The curves $\tau_s(\phi)$  also depend 
 on the spin quantization axis.  
 The effective spin-orbit coupling Hamiltonian\cite{Min06} for ideal flat graphene is 
 proportional to  the product of the  spin and valley  operators. 
 Therefore, it is not surprising that the spin relaxation 
 is different for $\hat{n}$ in plane and off-plane. We have verified that 120 
 degree rotations in the plane leave the spin lifetime unchanged, unlike the 90 
 degree rotation necessary to go from $\hat{n}=\hat{x}$ to $\hat{n}=\hat{y}$ (Figs.4a and b).    
 In general we find that spin lifetimes are a factor of 2 to 3 {\em longer} for spin 
 quantization axis in the plane than off-plane.  In  a  spin injection experiment 
 $\hat{n}$ would 
 be fixed by the magnetization orientation of the spin injector.  
 Present experimental results show the opposite trend\cite{Tombros08} 
 which is a further indication that other extrinsic mechanisms are dominant.

The curves in Figs.   \ref{anis-1}a,b and c are asymmetryc 
in the sense that $\int_{-\pi}^{0}\tau(\phi)d\phi\neq \int_{0}^{\pi}\tau_s(\phi)d\phi$, 
where $\tau_s$ is computed for a given valley.  
Interestingly, the symmetry is restored when summing over the two valleys, as shown 
in Fig. \ref{anis-1}d. In particular, 
we find the interesting relation: 
\begin{equation}
\tau_s(\phi,K)=\tau_s(-\phi,K').
\end{equation}

Altogether, the results of Fig. \ref{anis-1} show that the spin 
relaxation time due to scattering with flexural distortions is anisotropic on 3 counts: 
spin quantization axis, valley, and momentum direction.   Future work will determine if 
the preferred drift along a given direction, determined by an in-plane electric field, 
together with an externally imposed spin polarization, can serve to  generate an 
imbalance in the valley occupations, and thereby an 
orbital magnetization in graphene\cite{Niu2007}.

\section{Concluding remarks.}
In summary, we have shown that corrugations,
 that are ubiquitous in graphene in the form of dynamical flexural
 phonons or static ripples, 
enable a direct spin-flip mechanism 
to linear order in both the flexural field and in the spin-orbit coupling.
This mechanism 
provides an unavoidable   source of spin
relaxation that will set the upper limit for spin lifetimes once the extrinsic sources of spin relaxation that
prevail in state of the art experiments are removed.
 Such limit is however non-universal, as its
precise value depends on graphene's mechanical environment,
that determines the long-wavelength behavior of the flexural field.
At room temperature, intrinsic spin
  lifetimes in the microsecond  range are
  expected in a very wide range of situations.   Importantly, the intrinsic spin 
  relaxation time of  electrons in graphene shows a marked
dependence  on their momentum direction, valley and spin quantization axis.

Finally, whereas the existence of an  upper limit for
spin-lifetimes in graphene might present in the future an obstacle for certain
applications such as spin transistors, the  intrinsic spin-lattice
coupling could open the way for hybrid devices, where a confined
vibrational phonon could be coupled resonantly  to the
spin-flip transitions of Zeeman split confined Dirac
electrons. Microwave pumping of such system could result in a maser
behavior  of the phonon mode \cite{Roukes}.

\section*{Acknowledgments}  

This work has been financially supported by MEC-Spain (Grant Nos. FIS2010-21883-C02-01, FIS2009-08744,  and CONSOLIDER CSD2007-0010), European Union as well as Generalitat Valenciana, grant Prometeo 2012-11. T



\begin{thebibliography}{}
\bibitem{dorbitals}
In the calculations we have left out
$d$ orbitals, for simplicity. It has been shown that they  give the
dominant contribution to the SOC gap at the Dirac
points\cite{Fabian2009b,Fabian2010}, but they do not renormalize the
change of the gap with the flexural distortion\cite{Fabian2009b},
which is related to the spin-phonon coupling proposed here. In any
event, our results provide a lower limit for the instrinsic spin
relaxation mechanism, given the fact that including the  $d$ orbitals
will enhance  the effect.
\bibitem{kf}
This result is strictly valid for $k_F \gg k_BT/v_F$
which ensures that spin 
relaxation arises from a narrow interval around $k_F$.
\end{thebibliography}
\end{document}